\documentclass[prl,twocolumn,,aps,showpacs]{revtex4} 

\usepackage{graphicx}
\usepackage{dcolumn}
\usepackage{bm}
\usepackage{amssymb}
\usepackage{url}
\PassOptionsToPackage{hyphens}{url}\usepackage{hyperref}
\usepackage{color}

\begin{document}

\title{Election forensic analysis of the Turkish Constitutional Referendum 2017}

\author{Peter Klimek$^{1,2,*}$, Ra\'ul Jim\'enez$^3$, Manuel Hidalgo$^4$, Abraham Hinteregger$^1$, Stefan Thurner$^{1,2,5,6}$}
\affiliation{$^1$Section for Science of Complex Systems, CeMSIIS, Medical University of Vienna, Spitalgasse 23, A-1090, Austria\\
$^2$Complexity Science Hub Vienna, Josefst\"adter Strasse 39, A-1080 Vienna, Austria\\
$^3$Department of Statistics, Universidad Carlos III de Madrid, Spain\\
$^4$Department of Social Sciences, Universidad Carlos III de Madrid, Spain\\
$^5$Santa Fe Institute, 1399 Hyde Park Road, Santa Fe, NM 85701, USA\\
$^6$IIASA, Schlossplatz 1, A-2361 Laxenburg, Austria\\
$^*$\href{mailto:peter.klimek@meduniwien.ac.at}{peter.klimek@meduniwien.ac.at} }
\date{\today}

\begin{abstract}
With a majority of `Yes' votes in the Constitutional Referendum of 2017, Turkey continues its transition from democracy to autocracy.
By the will of the Turkish people, this referendum transferred practically all executive power to president Erdo\u{g}an.
However, the referendum was confronted with a substantial number of allegations of electoral misconducts and irregularities, 
ranging from state coercion of `No' supporters to the controversial validity of unstamped ballots.
In this note we report the results of an election forensic analysis of the 2017 referendum to clarify to what extent 
these voting irregularities were present and if they were able to influence the outcome of the referendum. 
We specifically apply novel statistical forensics tests to further identify the specific nature of electoral malpractices.
In particular, we test whether the data contains fingerprints for ballot-stuffing (submission of multiple ballots per person during the vote) 
and voter rigging (coercion and intimidation of voters).
Additionally, we perform tests to identify numerical anomalies in the election results.
We find systematic and highly significant support for the presence of both, ballot-stuffing and voter rigging. 
In 6\% of stations we find signs for ballot-stuffing with an error (probability of ballot-stuffing not happening) of 0.15\% (3 sigma event).
The influence of these vote distortions were large enough to tip the overall balance from `No' to a majority of `Yes' votes.
\end{abstract}


\maketitle

In 1996, Recep Tayyip Erdo\u{g}an, then-mayor of Istanbul, remarked that democracy can be compared with a bus ride, ``once I reach my stop, I get off'' \cite{Milliyet96}.
It seems that he arrived at one of these stops on April 16, 2017, when Turkish people went to the polls to vote on a constitutional reform package that among others would replace Turkey's parliamentary system with a presidential one.
The `Yes' won by a slight margin  — 51.4\% to 48.6\% or 1.38 million votes.
The narrow victory has been questioned by opposition forces alleging voting irregularities and even electoral fraud \cite{WSJ17, Observer17}. 
Some videos circulated of alleged electoral malpractices on the day of the election \cite{Observer17}.
There were reports on unverified (i.e. unstamped) ballots being cast, state coercion of `No' supporters, and election observers being kept from polling places \cite{WSJ17}.
The OSCE/ODIHR election observers noted that the referendum took place on an ``unlevel playing field'' and that ``observers were impeded in their observation during opening and voting'' \cite{OSCE}.
Further, there were reported cases of police presence at polling stations, police checking voter identification before granting access, as well as significant changes in the ballot validity criteria, effectively ``undermining an important safeguard and contradicting the law'' \cite{OSCE}.
Note that until the 2017 referendum, there were no indications that electoral fraud was a major concern in Turkish elections \cite{Esen16, Kemahl15}.

For a timely identification of electoral misconduct and to enable more targeted and efficient election observation missions, the newly emerging field of election forensics seeks to diagnose -- on a fully quantitative and data-driven basis -- to which extent a given type of malpractice might have impacted the outcome of an election \cite{Levin12}.
Often these tests focus on a disproportionate abundance of round numbers 
in the election results \cite{Cantu11,Berber12} (reflecting the human tendency to choose round numbers when making up the results) or the over-representation of certain digits in the results, i.e. violations of Benford's Law \cite{Mebane08, Pericchi11}.
Another type of election forensic tests aims at identifying irregularities in the distributions of vote and turnout numbers across polling stations, as well as correlations between these distributions \cite{Myagkov09, Levin09, J, Klimek12, Jimenez17}.
These statistical tools are often complemented by analyses of secondary data, such as exit polls or survey- and sampling data \cite{Prado11, Hausmann11}.

In this note we analyze the election results of the 2017 Turkish constitutional referendum by using different election forensics tools that have been recently proposed. We first test the data for the systematic occurrence of ballot-stuffing, i.e. the unlawful addition of a substantial numbers of ballots for a given party \cite{Klimek12}.
We then perform a test for the occurrence of the systematic coercion and intimidation of voters, i.e. a test for voter rigging \cite{Jimenez17}.
Finally, we carry out two additional tests for statistical irregularities, one based on the Second-Digit Benford's Law \cite{Pericchi11} and one based on the detection of outlier support \cite{Jimenez11}.

\section{Election data}

The election data were downloaded from the official website of the Turkish election commission \footnote{\url{https://sonuc.ysk.gov.tr}}.
We only considered results from Turkey itself, and did not include election results from polling stations in other countries, because the population eligible for voting was not clearly defined outside of Turkey.
In total, we analyze data for 166,679 polling stations grouped in 28,447 neighborhoods (or villages), belonging to 1,057 different districts, which are part of 81 provinces.
For each polling station $i$, we extracted the number of voters, $N_i$, the number of valid votes or turnout, $T_i$, 
as well as the number of `Yes' votes, $V_i$.
From these we obtained the relative turnout in percent, $t_i=T_i/N_i$, and the vote percentage, $v_i=V_i/T_i$.
Descriptive statistics of the polling stations are shown in table \ref{desc}, where we present mean values and standard deviations for $N_i$, $T_i$, $V_i$, $t_i$, and $v_i$.
Here and in the following analysis we excluded polling stations with an electorate of less than hundred people to rule out that our results are driven by such outliers.
It is important to stress that the concrete placing of the threshold does not alter the results. 
Almost identical results are obtained by placing the threshold at 0, 50 or 200.
About 1.3\% of all votes are not considered by implementing the threshold of 100.

\begin{figure}[tbp]
\begin{center}
 \includegraphics[width = 0.45\textwidth, keepaspectratio = true]{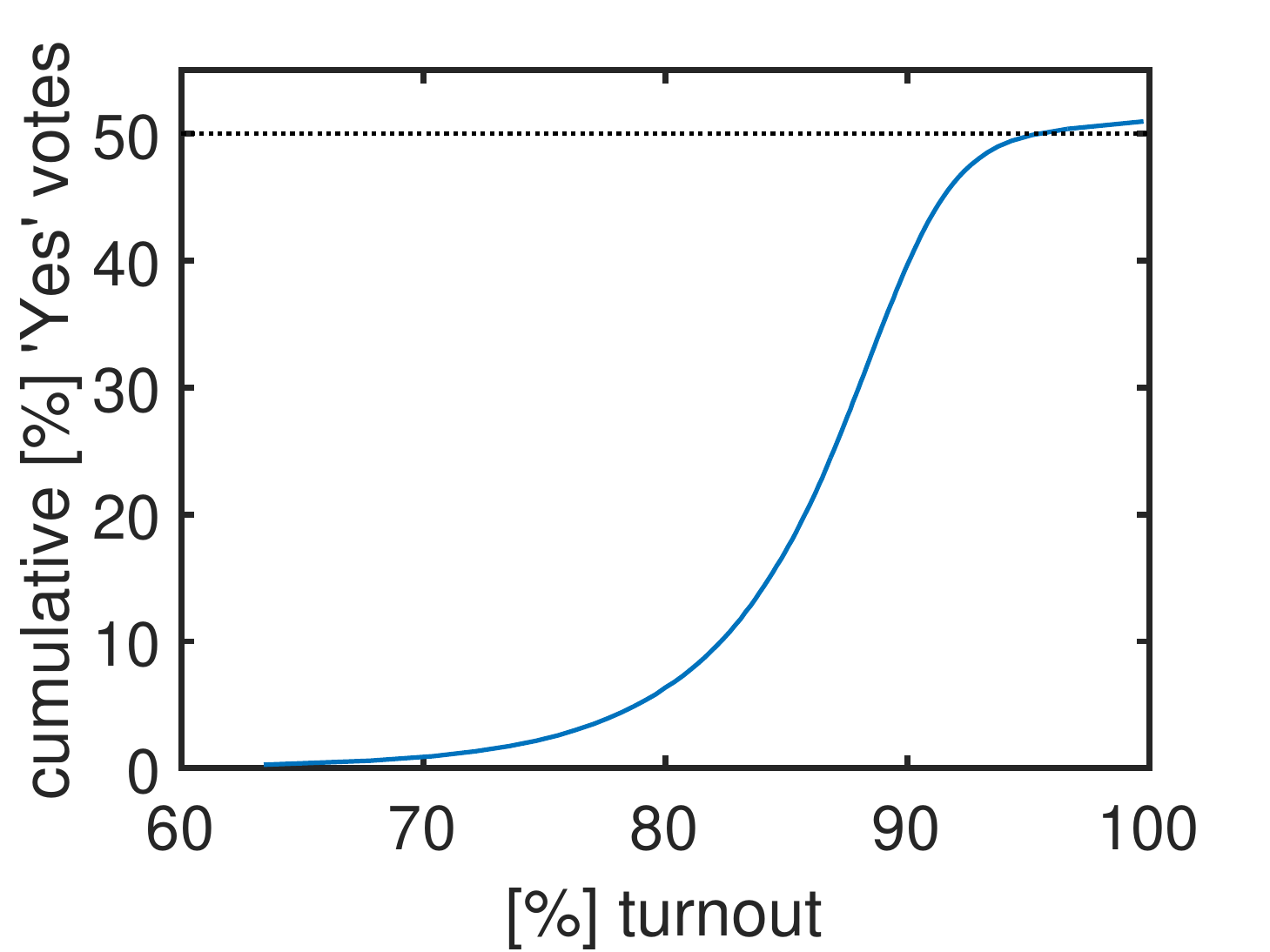}
\end{center}
 \caption{For a given level of turnout, the cumulative vote percentage of stations with this level or lower is shown. A majority of more than 50\% is achieved with the inclusion of high-turnout stations.}
 \label{hists}
\end{figure}

The cumulative number of `Yes' votes is shown as a function of turnout in figure \ref{hists}. 
For each level of turnout shown on the $x$-axis, the total number of votes from stations with this level or lower is shown on the $y$-axis.
The vote percentages cross the 50\% threshold with the inclusion of polling stations with a turnout of close to 100\%.

\begin{table}[ht]
\caption{Descriptive statistics of polling stations in the 2017 Turkish constitutional referendum.}
\begin{tabular}{l | l | l}
variable $x_i$ & mean $\langle x_i \rangle$ & standard deviation $\sigma (x_i)$  \\ 
\hline
number of voters, $N_i$  & 332 & 109 \\
turnout, $T_i$  &  285 & 86 \\
Yes' votes, $V_i$ & 146 & 74 \\ 
relative turnout, $t_i$  &  0.86 & 0.085 \\
vote percentage, $v_i$ & 0.53 & 0.23 \\ 
\end{tabular}
\label{desc}
\end{table}

\section{Ballot-stuffing test}  

\begin{figure*}[tbp]
\begin{center}
 \includegraphics[width = 0.9\textwidth, keepaspectratio = true]{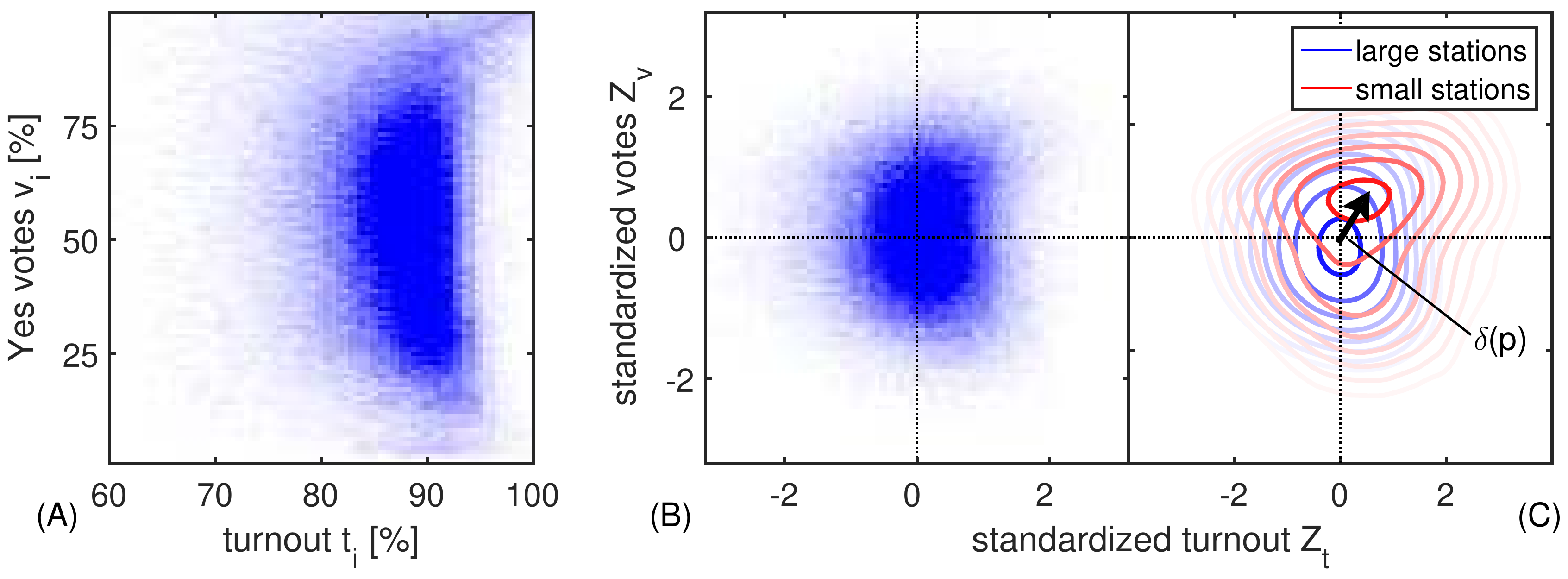}
\end{center}
 \caption{Election forensic fingerprints for the Turkish constitutional referendum 2017. (A) The fingerprint shows the joint vote-turnout distribution where the color intensity indicates the number of stations with a given vote and turnout. The distribution is smeared out towards high vote and high turnout numbers, which is characteristic for ballot-stuffing. (B) Standardized fingerprints as defined in the text; they can be used to adjust for geographic heterogeneities in the data. (C) Traces of voter rigging can be identified by comparing the standardized fingerprints of small (red lines) and large (blue lines) polling stations. Small stations are particularly susceptible to voter coercion and intimidation, which results in their displacement toward inflated votes and turnout (shift of small stations shown as red lines toward the upper right corner).}
 \label{fp}
\end{figure*}

It has been shown that specific types of electoral fraud introduce 
odd correlations between turnout and vote numbers that can not be accounted for by demographic or geographic characteristics 
\cite{Klimek12}.
The presence of such correlations can be estimated by using so-called {\em election fingerprints}, i.e. the joint vote-turnout distribution that can be represented in 2-d histograms \cite{Klimek12}. The fingerprint for the Turkish 2017 referendum is shown in figure \ref{fp}(A), where the color intensity (blue) is proportional to the number of polling stations with the corresponding percentage of votes ($x$-axis) and turnout ($y$-axis).
In the absence of non-linear vote-turnout correlations, the bulk of the distribution in figure \ref{fp}(A) should show a circular or elliptical 
symmetry.
The occurrence of ballot-stuffing in a district would inflate the turnout and at the same time increase the vote percentages.
If this happens in a substantial number of polling stations, the vote and turnout numbers become correlated and the elliptical 
symmetry in the fingerprints is broken.
For the Turkey 2017 data we observe a bulk that is spread out particularly along the vote dimension, but is rather narrow in turnouts.
For high votes and high turnout, this bulk is clearly smeared out towards the upper right corner of the plot, which is fully consistent with a ballot-stuffing scenario.

To assess whether the deviations observed in the fingerprint are indeed statistically significant traces of ballot-stuffing, we apply the parametric model that was proposed in \cite{Klimek12}. In a nutshell, this model assumes a fingerprint with normally distributed and independent vote- and turnout numbers, with means and standard deviations that were estimated from the data. The model tests if the skew towards higher numbers in the observed vote distribution coincides with a similar skew in the observed turnout distribution, as would be characteristic for ballot-stuffing.
The result is a fraud parameter, $f_i$, which represents the fraction of polling stations with ballot-stuffing-like distortions in their respective vote and turnout numbers \footnote{Note that the parametric model proposed by Klimek et al. also considers a different, extreme type of ballot-stuffing where vote and turnout numbers are inflated to 100\%, each. In the present analysis, there are no indications of such extreme statistical irregularities.}
For the Turkey 2017 data we obtain a non-zero fraud parameter, 
\begin{equation}
f_i = 0.058 \pm 0.019 \quad. 
\label{fi}
\end{equation}
Note that this is roughly a 3 sigma effect, meaning that the mean of the distribution of $f_i$ is three standard deviations from the assumption of no ballot-stuffing, which is $f_i = 0$. 
We find a shape parameter of $\alpha=1.3 \pm 0.2$. 
The shape parameter measures to which extent the ballot-stuffing process in the parametric model is combined with a deliberate wrong-counting or recasting of ballots.
A shape parameter larger than one indicates that ballot-stuffing dominates over the wrong-counting process. 
This means there is a highly significant effect in the Turkish election fingerprint that is compatible with the ballot-stuffing hypothesis. 
Compared with the irregularities observed in recent Russian elections, these deviations are relatively weak but nevertheless systematic and statistically significant.

\section{Voter rigging test}

\begin{figure}[tbp]
\begin{center}
 \includegraphics[width = 0.45\textwidth, keepaspectratio = true]{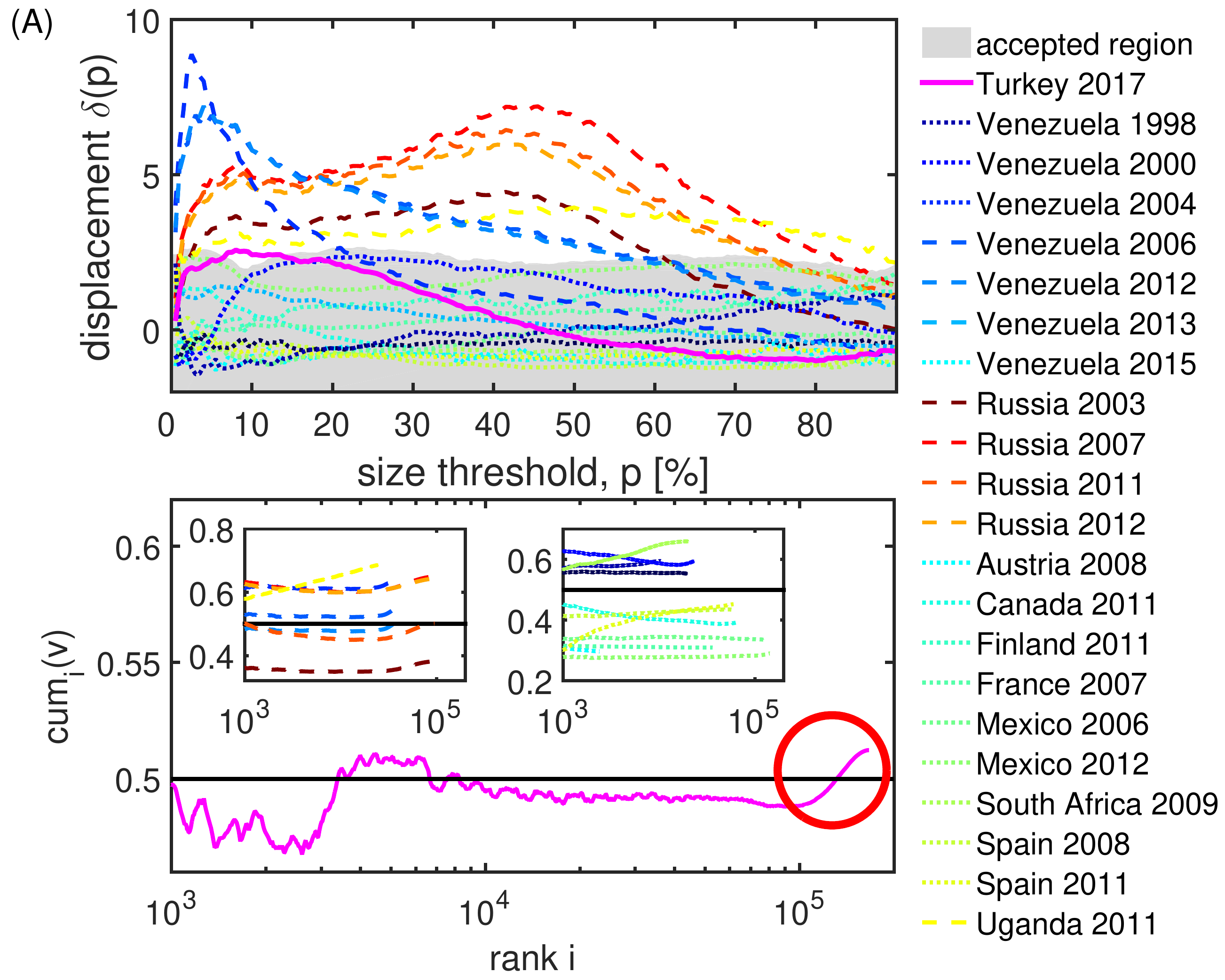}
\end{center}
 \caption{Results for the statistical test for voter rigging. (A) An accepted region for the displacements is constructed from the confidence interval of displacements observed in the reference set of trustworthy elections. There is a significant displacement $\delta(p)$ between small and large polling stations with values that lie outside this accepted region for Turkey 2017 (full, magenta line). The displacement sizes are substantially smaller than those observed in Russian or recent Venezuelan elections (shown as blue and red dashed lines). Reference elections are shown as dotted lines. (B) We rank all stations in Turkey by their size and show the cumulative vote percentage $cum_i(v)$ computed over all stations with a size larger than the given rank. For higher ranks $i$, an increasing number of small stations is included. It is the addition of small units with inflated votes and turnouts that pushes the results over the 50\% line and leads to a majority of Yes' votes (highlighted by a red circle). In the insets, we show the same relationship for other elections that (left) show significant displacements or (right) belong to the set of reference elections.}
 \label{vr}
\end{figure}

In some cases irregularities in the fingerprints can be explained by geographic heterogeneities, for instance due to different mobilization effects across urban and rural areas.
A way to account for such natural correlations in the data is to compare each polling station to other stations that are in close geographic proximity \cite{Jimenez17}.
In the case of Turkey 2017, we compared the vote and turnout numbers of each station to the average values that have been observed in the same district.
For a polling station $i$ in district $A$, we defined the electoral neighborhood, $Nb(i)$, as all other polling stations in $A$.
The standardized vote percentage of station $i$, $Z_v(i)$, is then given by the $Z$ score, 
\begin{equation}
Z_v(i) = \frac{v_i-\mu_{j \in Nb(i)}(v_j)}{\sigma_{j \in Nb(i)} (v_j)} \quad,
\label{Zv}
\end{equation}
where $\mu_{j \in Nb(i)}(v_j)$ and $\sigma_{j \in Nb(i)} (v_j)$ denote the mean and standard deviation over all districts in the electoral neighborhood of $i$, respectively.
The standardized relative turnouts are, 
\begin{equation}
Z_t(i) = \frac{t_i-\mu_{j \in Nb(i)}(t_j)}{\sigma_{j \in Nb(i)} (t_j)} \quad.
\label{Zt}
\end{equation}
The so-called standardized fingerprint (2-d histogram of the standardized vote and turnout numbers, $Z_v$ and $Z_t$) is shown in figure \ref{fp}(B), \cite{Jimenez17}.
Using this representation, it becomes possible to address the issue of of voter rigging.
The key hypothesis in this test is that smaller polling stations are more susceptible to coercion and intimidation of voters, since (i) it is easier to identify opposing individuals, (ii) there are fewer eyewitnesses, and (iii) such stations are visited less frequently by election observers.
Consequently, voter rigging shows up in the standardized fingerprint by a displacement (a shift towards higher vote and higher turnout numbers, upper right corner) of the fingerprint of small stations away from the fingerprint of large stations.
Small stations were defined as those with an electorate size, $N_i$, that is located in the lowest $p$-th percentile of all electorate sizes.
In figure \ref{fp}(C) we show the standardized fingerprints in the form of ``iso-density'' lines for small (red) and large (blue) polling stations for $p=$10\%.
The size of the displacement generally depends on the size threshold $p$ and is denoted by $\delta (p)$, see arrow in figure \ref{fp}(C). 
It is apparent that the fingerprints for small stations are obviously shifted towards the upper right  corner of the figure, as would be expected from voter rigging.
As for the ballot-stuffing case, a visual examination of the (standardized) fingerprints alone is not conclusive and 
a hypothesis test \cite{Jimenez17} is needed to assess whether the observed displacement between small and large polling stations is statistically significant and indeed consistent with voter rigging.
The idea behind the test of Jimenez et al. \cite{Jimenez17} is to construct a baseline for expected displacements between small and large stations based on a reference set of trustworthy elections. From these elections a region of an ``acceptable displacement size''  is derived.  
The acceptable displacement size was obtained from an analysis of 21 different elections in ten countries. 
For a detailed description of the test and the data used, see \cite{Jimenez17}.
Given the acceptable region, for a given election, one can now check if the actually observed displacement between small and large stations for a range of size thresholds $p$, falls within this region. 
If the displacement is larger than the 95\% confidence interval of displacements observed in the reference set, this signals statistical significance at the 5\% level. 
Here, we extend this analysis to the data of the Turkish 2017 constitutional referendum; results are shown in figure \ref{vr}.
In figure \ref{vr}(A) we show the average displacement, $\delta(p)$ between small and large stations in the standardized fingerprint as a function of $p$ for the extended dataset, including the Turkey 2017 referendum.
For small size thresholds $p$, the Turkish data shows indeed larger displacements than allowed for in the acceptable region. This indicates 
statistically significant signs of voter rigging.
Compared to the Russia and Venezuela, the signatures for voter rigging in Turkey are smaller, as it was in the ballot-stuffing test.
Finally we estimate the impact of the voter rigging effect in the data. For this purpose we 
first rank each polling station by its electorate size in decreasing order, and then 
compute the cumulative vote percentages, $cum_i(v) = \sum_{j<i}^i  V_i/\sum_{j<i}^i  T_i $, over all stations with a rank $j$ less than $i$. 
In figure \ref{vr}(B) we present $cum_i(v)$ as a function of the rank. 
The signal for voter rigging can be seen in the high rank region (small stations) of the cumulative vote percentage curves, 
where a sharp increase for the smallest stations is seen (circle). This signal for voter rigging is a typical pattern that was also 
found in Russia and Venezuela, see inset 1. In elections where no fraudulent actions were reported, these patterns are missing, see inset 2.
For the Turkish constitutional referendum we see that the cumulative effect of the distortions in small stations was enough to tip the results toward a majority of `Yes' votes. 
Only with these distortions the cumulative votes cross the  50\% line in figure \ref{vr} B. 

The voter rigging test also allows us to identify which provinces in Turkey contributed most to the observed irregularities.
For this we computed the displacements, $\delta (p)$, for each of the 81 provinces separately, treating each province as if it were an individual country.
One can then average $\delta (p)$ over all used values of $0<p<90$ to obtain a single number for the importance of the voter rigging effects in each province.
The ten provinces with the strongest effects of voter rigging in decreasing order are 
 \c{S}anliurfa,  K\"utahya, Bayburt, D\"uzce, K\'il\'is, \c{C}ankiri, G\"um\"u\c{s}hane, Bolu, Kastamonu, Tokat with respective average of $\delta (p)$ around two and maximal displacements that range from 2.9 to 4.3.
These provinces are spread more or less equally over the entire country but tend to have a low population density (i.e. four of the above provinces rank among the ten least populated provinces, whereas the most populated one is D\"uzce at rank 15 of 81).

\section{Further tests for statistical irregularities}

The Benford test for the second significant digit is one of the most commonly used tools in election forensics.
Benford's Law states that the second significant digit of the number of votes, $V_i$, must be a random number with a certain, specified frequency distribution, namely a so-called power law \cite{Pericchi11}.
Deviations from this phenomenological law might indicate an influence of human thought (such as rounding or cutting off certain numbers), however, it is not at all clear whether such deviations can be related to concrete forms of electoral fraud \cite{Deckert10}.
Here, we tested for Benford's Law by following the protocol proposed in \cite{rjmh14}.
Therefore, we consider only electoral units with three significant digits for testing the null hypothesis $H_0$: The data is consistent with the Benford's Law for the second significant digit. 
As a measure of being right when we assert $H_0$ is true, we compute 
the Bayesian posterior probability proposed by Pericchi and Torres \cite{Pericchi11}, denoted by $P(H_0|\mbox{data})$.
At the finest aggregation level (polling stations), we observe a large deviation from the law, with $P(H_0|\mbox{data}) < 10^{-120}$.
As far as we know, this discrepancy has been observed only in scenarios of manipulating electoral data.
We repeat the analysis for the next data aggregation level (villages).
They group in average over 3 poling stations. 
Even at this aggregation level, the distributions deviate significantly from Benford's Law, with $P(H_0|\mbox{data}) < 10^{-10}$.
In all cases considered so far, aggregated data distributed on such an order of magnitude confirmed Benford's Law \cite{rjmh14}.
The significant deviations found in Turkey constitute therefore a highly irregular observation, which was not even found in Venezuelan data.

Another statistical test for irregularities in election data is based on the assumption that voters are assigned to polling stations in their 
villages in a way that should {\it not} depend on their voting behavior.
By randomly permuting the way how voters (as inferred from the data) are assigned to polling stations in their village, a null model can be formulated for the (non-)randomness of the assignment of voters to their polling stations \cite{rjmh14}.
Following the test procedure described in \cite{rjmh14}, we found that the standardized differences between a random and the actual voter assignment were indeed systematically out of the 99\% normal confidence interval. 
Until now, such extreme deviations have only been observed in cases that where accompanied by a substantial amount of fraud claims \cite{rjmh14}.

\section{Conclusions}

In this note we reported the results of an election forensic analysis of the Turkish constitutional referendum in 2017.
We applied several recently proposed statistical tests to test for the elementary and low-tech mechanisms of election fraud, ballot-stuffing and voter rigging, respectively.
For both we find systematic and statistically significant indications in the data.
In particular our analysis suggests the existence of ballot-stuffing in about 6\% of the polling stations and a combined effect with voter rigging that was just large enough to change the outcome of the referendum from `No' to `Yes'.
The official report of the election observers \cite{OSCE} criticized (i) the validity of unstamped and unverified ballots during the election and (ii) police presence at polling stations to check voter identifications before granting access.
The reported large-scale addition of unverified ballots would clearly result in a positive test for ballot-stuffing, whereas voter intimidation at the polling stations would show up as a positive test for voter rigging. 
In this sense, the mere presence of these types of electoral malpractices in the Turkish referendum is not a new result {\em per se}.
However, our analysis shows for the first time in a quantitative and data-driven way that the impact of these irregularities on the election outcome was decisive in transforming Turkey into an executive presidency.


\begin{thebibliography}{99}

\bibitem{Milliyet96} ``Ergo\u{g}an'dan anayasa yorumu'', Milliyet July 14 1996, \url{http://gazetearsivi.milliyet.com.tr/Arsiv/1996/07/14}.

\bibitem{WSJ17} ``Inside Turkey's Irregular Referendum'', The Wall Street Journal April 25 2017, \url{https://www.wsj.com/articles/inside-turkeys-irregular-referendum-1493150990}.

\bibitem{Observer17} ``Turkey: Videos show electoral fraud and ballot stuffing'', The Observers, France 24, April 18 2017,  \url{http://observers.france24.com/en/20170418-turkey-videos-show-electoral-fraud-ballot-stuffing}

\bibitem{OSCE} ``Republic of Turkey Constitutional Referendum 16 April 2017. OSCE/ODIHR Limited Referendum Observation Mission Final Report''. OSCE/ODIHR 22 June 2017, Warsaw, Poland.

\bibitem{Esen16} Esen B and Gumuscu S (2016) Rising Competitive Authoritarianism in Turkey. {\it Third World Quarterly} {\bf 37}: 1581-1606. 

\bibitem{Kemahl15} Kemahlio\u{g}lu \"O (2015) Winds of Change? The June 2015 Parliamentary Election in Turkey. {\it South European Society and Politics} {\bf 20}: 445-464.

\bibitem{Levin12} Levin I, Cohn GA, Ordeshook PC, and Alvarez RM (2009) 
Detecting voter fraud in an electronic voting context: An analysis of the unlimited reelection vote in Venezuela,
in EVT/WOTE'09 Proceedings of the 2009 conference on Electronic voting technology/workshop on trustworthy elections, USENIX Association, Berkeley.

\bibitem{Cantu11} Cant\'u F and Saiegh SM (2011) 
Fraudulent Democracy? An Analysis of Argentina's Infamous Decade Using Supervised Machine Learning. 
{\em  Political Analysis} {\bf 19}: 409--433.

\bibitem{Berber12} Berber, B and Sacco, A (2012)
What the Numbers Say: A Digit-Based Test for Election Fraud. 
{\em  Political Analysis} {\bf 20}:211--234.

\bibitem{Mebane08} Mebane W (2008)
Election forensics: The second-digit Benford's law test and recent American presidential elections,
in {\em Election Fraud: Detecting and Deterring Electoral Manipulation}, eds Alvarez RM, Hall TE and Hyde SD (Brooking Press, Washington DC), pp 162--181. 

\bibitem{Pericchi11} Pericchi L and Torres D (2011)
Quick anomaly detection by the Newcomb-Benford Law, with applications to electoral processes data from the USA, Puerto Rico, and Venezuela.
{\em  Statist Sci} {\bf 26}:513--527. 

\bibitem{Myagkov09}
Myakgov M, Ordeshook PC, and Shaikin D (2009)
{\em The Forensics of Election Fraud}, Cambridge University Press. 

\bibitem{Levin09}
Levin I, Cohn GA, Ordeshook PC, and Alvarez RM (2009) 
Detecting voter fraud in an electronic voting context: An analysis of the unlimited reelection vote in Venezuela,
in EVT/WOTE'09 Proceedings of the 2009 conference on Electronic voting technology/workshop on trustworthy elections, USENIX Association, Berkeley.

\bibitem{Jimenez11}
Jim\'enez R (2011) 
Forensic analysis of the Venezuelan recall referendum. 
{\em Statist Sci}{\bf 26}:564--583.

\bibitem{Klimek12}
Klimek P, Yegorov Y, Hanel R, and Thurner S (2012)
Statistical detection of systematic election irregularities.
{\em Proc Natl Acad Sci USA}{\bf 109}:16469--16473.

\bibitem{Jimenez17}
Jim\'enez R, Hidalgo M, Klimek P (2017)
Testing for voter rigging in small polling stations
{\it Science Advances} {\bf 3}:e1602363.

\bibitem{Prado11}
Prado R and Sans\'o B (2011)
The 2004 Venezuelan presidential recall referendum: Discrepancies between two exit polls and official results.
{\em Statist Sci} {\bf 26}:502--512.

\bibitem{Hausmann11}
Hausmann R and Rigob\'on R (2011) 
In search of the black swan: Analysis of the statistical evidence of fraud in Venezuela. 
{\em Statist Sci} {\bf 26}:543--563.

\bibitem{Deckert10}
Deckert J, Myagkov M, Ordeshook PC (2010)
The irrelevance of Benford's Law for detecting fraud in elections.
CalTech/MIT Voting Technology Working Paper No 9.

\bibitem{rjmh14}
Jim\'enez R and Hidalgo M (2014)
Forensic analysis of Venezuelan elections during the Ch\'avez presidency.
{\em PLoS ONE} 9(6):e100884.

\end{thebibliography}
\end{document}